\documentclass[preprint,preprintnumbers,amsmath,amssymb]{revtex4}
\usepackage{psfig}

\newcommand{\beq}{\begin{eqnarray}}
\newcommand{\eeq}{\end{eqnarray}}
\newcommand{\vve}{\vec{v}_{ext}}
\newcommand{\vvi}{\vec{v}_r}
\newcommand{\vq}{\vec{q}}
\newcommand{\vro}{\vec{r}_0}

\newcommand{\go}{\omega}
\newcommand{\gO}{\Omega}
\newcommand{\divg}{\vec{\nabla}\cdot}
\newcommand{\grad}{\vec{\nabla}}
\newcommand{\D}{{\cal D}}

\begin{document}
\begin{center}
{\bf\LARGE Shape Fluctuations in Randomly Stirred Dilute Emulsions}\\\vspace{.7cm}
{\bf Gad Frenkel}\\
{\em Department of Earth Science and Engineering, Imperial College, London SW7 2AZ, United Kingdom}
\end{center}

\begin{abstract}
In this paper we consider the effects of the interaction between droplets or other deformable objects in an emulsion under random stirring of the host fluid. Our main interest is to obtain autocorrelation functions of the shape fluctuations in such randomly stirred host fluids, beyond the dilute limit regime. Thus, a system of deformable objects immersed in a host liquid that is randomly stirred is considered, where the objects interact with each other via the host liquid. Keeping expressions in the first order in the density of objects and in deviation of objects shapes from spherical, the shape of each object is expanded in spherical harmonic modes and the correlations of these modes are derived. The special case of objects that are governed by surface tension is investigated. The interaction between objects is explicitly formulated and the deformation correlations are obtained.

\end{abstract}

%
\section{INTRODUCTION}
%

Single deformable objects such as droplets of one liquid immersed in another liquid fluctuate in shape in response to random external stirring \cite{gady3}. 
The purpose of this paper is to investigate the effect of interaction between such deformable objects \cite{schwartz90b} on the way they deform due to fluctuations of the velocity field in the fluid in which they are immersed.
Numerous authors have studied the fluctuations and diffusion of single
deformable objects due to thermal agitation
\cite{sparling89,safran87,schwartz91,lisy94,gang,foltin,zilman,gady2}.
Nevertheless, there are
clearly other ways in which systems are agitated. In industrial
and biological environments, the host liquid is often stirred,
shaken or pumped in ways which are very different from thermal agitation.
The list of examples is not restricted to
artificial processes. It also includes natural processes such as
Brownian motion of small beads induced by the collective motion of
bacteria \cite{libchaber} and nano-scale mechanical fluctuations
of the red blood cell surface that have been measured and shown to
depend strongly on the biochemical environment and not only on
temperature \cite{krol,levine,mittelman1,mittelman2}. For this reason the external velocity field agitating the system is taken to be more general than that corresponding just to thermal motion. The article provides thus the general equations describing the effect of a finite density of deformable objects on the diffusion and shape fluctuations of a single object, to linear order in the density.\\ 

%
The system considered has the following properties. \\
(a) The deformable objects are fluid, in the sense that the velocity field is well defined everywhere (both inside and outside the object). No slip and no penetration conditions are assumed at the interface of the deformable object. Hence, each surface element moves with the velocity of the flow at its position.
%
In addition, both the objects and the host fluid are incompressible.
%
The objects are characterized by an energy that depends on their
shape (i.e. changing the orientation or switching places of two
surface particles while keeping the shape constant does not change
the energy). The main example considered in this paper is surface tension \cite{Ta1,Ta2}. However the description can be extended to Helfrich bending energy \cite{helfrich76,lisy98} and other cases where the shape
of minimum energy is nearly spherical. Deformation of the shape changes
the energy, exerts a force density on the liquid and therefore
generates an additional velocity field, denoted by
$\vec{v}_{\psi}$.\\
(b) The hydrodynamic equations of the host liquid are
  linear in the velocity ( i.e. a velocity field induced by several
  sources is equal to the sum of the velocity fields induced by each source separately). 
  For instance, if the flow is governed by the
  Navier Stokes equation, then the linearity implies that the
  Reynolds number is small and that the Stokes approximation is
  applicable. The actual velocity field is the sum of the imposed
  velocity field , $\vec{v}_{ext}$ (the velocity field that would have
  existed if
  the objects were absent), the velocity field induced by the
  deformations of the object under consideration, $\vec{v}_{\psi}$, and $\vvi$ which is the velocity field created by the rest of the deformable objects,
\beq  \label{eq:VextPVpsi}
\vec{v} = \vec{v}_{ext} + \vec{v}_{\psi} + \vvi.
\eeq
(c) The external velocity field is assumed to be
  random. The correlations are assumed depend only on distance and time difference. 
Furthermore, the dependence on the time difference is taken to be extremely short ranged (Dirac $\delta$ function in the time difference). In principle, equations for the dependence of the shape correlations on the density of deformable objects can be worked out for any dependence of the velocity correlations on time. These are very complicated, however, and the above choice of the dependence of the external velocity correlations on time simplifies matters considerably and is certainly realistic in many cases. It is important to note that the results obtained here are not used to
  determine the external velocity correlations. Those correlations are just taken as a given input. For example, in the
  special case of thermal agitation
  the velocity correlation was calculated from first
  principles \cite{gady2} and only then
  used to calculate the diffusion constant and deformation characteristics of a deformable object immersed in the liquid.
  In addition the external velocity is assumed to be small
  enough to allow the body to remain almost spherical.

Since we assume small deviations from the spherical shape it is only natural to describe the surface shape of the objects using spherical harmonics.
Consider a spherical body which is moving and is slightly deformed. The equation
\beq   \label{eq:fOmegat}
\frac{\rho}{R} + f(\Omega,t) - 1 = 0
\eeq
defines its
surface, yielding for each spatial direction, $\Omega$, the distance,
$\rho \equiv |\vec{r} - \vec{r}_0|$, of the surface from the
centre of the body, $\vec{r}_0$. $R$ is the radius of the undeformed
sphere. The deformation function, $f(\Omega,t)$,
defines the shape and may be expanded in spherical harmonics,
$f(\Omega,t)=\sum_{l=1}^\infty \sum_{m=-l}^l
f_{lm}(t)Y_{lm}(\Omega)$
(clearly the $Y_{00}$ term can be absorbed in the definition of $R$).
The goal is to obtain
the correlations between the deformation coefficients, $f_{lm}(t)$.
The centre of the object, $\vec{r}_0$, is chosen to be the point
around which the deformation coefficients with $l=1$ vanish: $f_{1
m}=0$. A different definition of the centre will introduce three
different equations for the deformation coefficients with $l=1$.
These are not interesting, as far as the shape is concerned, since
in the first order of the deformation the spherical harmonics with $l=1$ describe a rigid translation of the object
\cite{schwartz88,gady-physica-A}.\\

The random velocity field and the effect of the
interaction between objects induce fluctuations in the values of the
deformation coefficients describing any of the objects. Consider the autocorrelation of the deformation
coefficient $f_{l,m}^i$ of the $i$'th object $\langle
f_{l,m}^i(\go) f_{l,-m}^i(-\tilde{\go})\rangle$, where $f_{lm}^i(\go)$ and $f_{lm}^i(\tilde{\go})$ represents the Fourier transforms (FT) of the deformation coefficients of the i'th object in respect to time.
The autocorrelation is expanded in orders of $n$. To first
order in $n$ it is given by
\beq
\left\langle f_{l,m}^i(\go) f_{l,-m}^i(-\tilde{\go})\right\rangle =
{\cal G}_{l,0}(\go,\tilde{\go}) +
{\cal G}_{l,1}(\go,\tilde{\go})\cdot n
\eeq
The first term on the right hand side of the equation (3) above gives the shape correlations of a single object , that has been described previously \cite{gady3} . The second term represents the correction to the shape correlations due to a small but finite density of deformable objects. The purpose of this article is to obtain that correction.

The paper is organized as follows. Section \ref{sec:single} deals with a single deformable object in random flow. The aim of this section is to introduce the basic definitions of flow and present the zero-order terms in the expansion of the shape correlations in the density of objects $n$. In section \ref{sec:flm} the first order terms of the shape correlation functions are derived. Section \ref{sec:white} deals with the special case of identical droplets that are governed by surface tension in a random flow. In order to improve readability, Parts of the derivation of the first order terms and the velocity field induced by a deformable body governed by surface tension are left to the appendix.

\section{A single object in random flow} \label{sec:single}
%
The response of a single object immersed in a host liquid to an external random flow has been described in previous work \cite{gady3}. The results are sketchily repeated here for the benefit of the reader as the general equations obtained here are to be exploited in the next section by replacing the external velocity field by the velocity field seen by the object when a finite density of deformable objects is immersed in the liquid. The latter velocity field is the sum of the imposed velocity field and the velocity fields induced by all the other deformable objects. 
The correlations of the deformation coefficients as well as diffusion constant of the centre will be obtained here in terms of the correlations of the external velocity. The diffusion constant will be needed in the next section to obtain the first order correction in the density of the shape correlations.\\
The no-slip and no-penetration conditions yield an equation of
evolution for the deformation coefficients \cite{schwartz88},
\beq \label{eq:basic}
\frac{\partial f_{lm}(t)}{\partial t} + \lambda_l f_{lm}(t) =
-Q_{lm}(t).
\eeq
The effect of the velocity field induced by the deformable object itself is represented by the second term on the left hand side above. 
The $\lambda_l$'s characterize the way in which a deformation with definite $l$ decays to zero in the absence of an external
velocity and other objects. Different physical systems are characterized by different
sets of $\lambda_l$ \cite{schwartz88,safran87,foltin,gang95}.
The term on the right hand side, $Q_{lm}$, is given by
\beq   \label{eq:Qlm}
Q_{lm}
= \frac{1}{R} \int d\Omega \left\{
    \hat{\gO}\cdot\left[\vec{v}_{ext} -
      \dot{\vec{r}}_0\right]Y^*_{l,m}(\Omega)\right\},
\eeq
where the external velocity field, $\vec{v}_{ext}$, is evaluated on
the undeformed body and $\hat{\gO}$ is a unit vector in the
direction of the spatial angle $\Omega$ (for further detail see
\cite{gady2}).
The velocity of the centre $\dot{\vec{r}}_0$ contributes only to
$Q_{1,m}$. In addition, the definition of the centre implies that
\beq
Q_{1,m}=0,
\eeq
for all $m$.\\
%
%
It is convenient to write the correlation function of the external
velocity field in momentum space. This is so because the random
velocity field is transversal when the fluid is
incompressible. Consequently, in real space, the flow must always be
correlated in a very complex way. On the other hand,
in momentum space, the transversal part of a general field is easily obtained:
\beq  \label{eq:vext}
v_{ext}^i(\vec{q}\ ) \equiv \sum_j \left( \delta_{ij}-\frac{q_i
    q_j}{q^2}\right) u^j(\vec{q}),
\eeq
where $\vec{u}$ is a general vector field and $i$ and $j$ denote
Cartesian components. The bracketed term is the
projection operator that removes the longitudinal part of
$\vec{u}$, and therefore yields a general transverse velocity field
$\vec{v}_{ext}$.
Next, the correlations of the external velocity are easily
expressed using the correlations of the general field $\vec{u}$.
\beq   
\label{eq:fcorelation}
\left\langle u^l(\vec{q},t_1) u^m(\vec{p},t_2) \right\rangle =
\delta_{lm}\delta(\vec{q}+\vec{p})\phi(q , |t_2-t_1|), 
\eeq 
where $\phi$ is a general function of $q$ and the time difference (with the only limitation that its Fourier transform in the time difference is non-negative).
As was mentioned before, this investigation is restricted to cases where the external velocity is uncorrelated
in time, 
\beq \label{eq:PhiWhite}
\phi(q,t)=\tilde{\phi}(q)\delta(t),
\eeq
and where the mean of the velocity field vanishes,  
$\left\langle
u^l(\vec{q},t)\right\rangle = 0$.
\\
%
Using the above, the diffusion coefficient of the centre of the
deformable object is obtained \cite{gady-physica-A},
\beq
D=\frac{8\pi}{3}\int_0^\infty q^2\ dq \ \tilde{\phi}(q){\cal J}^2(qR),
\eeq
where ${\cal J}(x)=j_0(x)+j_2(x)$ and $j_n(x)$ is the spherical Bessel
function of order $n$.\\
The correlations of the deformation coefficients, $f_{l,m}$, \cite{gady3} are
given by 
%
%
%
\beq
\label{eq:flmflmdelta} \left\langle  f_{lm}(t) f_{l'm'}(t+\Delta
t)\right\rangle_{t \rightarrow \infty} = {\bf Q}_{l l }
\frac{e^{-\lambda_{l}|\Delta t|}}{2 \lambda_l}  \ \delta_{l',l} \
\delta_{m',-m}, \eeq
where 
\beq   \label{eq:Q00} {\bf Q}_{l l'}
\equiv \frac{1}{R^2} \int d\Omega \int d\Omega' \int d^3q \
Y_{l0}^*(\Omega)Y_{l'0}^*(\Omega') \Big[ \nonumber \\
\hat{\Omega}_i \hat{\Omega}_j' e^{-i\vec{q}\cdot(\hat{r} -
\hat{r}')R} \left[\delta_{ij} - \frac{q_i q_j}{q^2}\right]
\tilde{\phi}(q) \Big] .
\eeq
The contribution to the autocorrelation is presented here for completeness after being Fourier transformed in time,
\beq \label{eq:flmG0}
{\cal G}_{l,0}(\go,\tilde{\go})
=\frac{\delta(\go -\tilde{\go})}{4\pi^2(\lambda_l^2+\go^2)}
 {\bf Q}_{l l}.
\eeq
%


\section{First Order Correction} \label{sec:flm}
The aim of this section is to obtain the correction to the autocorrelation of the object's shape, which is linear in the density of objects $n$.
%
%
The full autocorrelation of deformation coefficients is obtained by replacing $\vec{v}_{ext}$ in equation (\ref{eq:basic}) by the sum of the external velocity and the velocity induced by the other objects,
\beq \label{eq:37}
\left\langle f_{l,m}^i(\omega) f_{l,-m}^i(-\tilde{\omega})\right\rangle
= \frac{1}{\lambda_l +i\omega}\frac{1}{\lambda_l
  -i\tilde{\omega}}\frac{1}{(2\pi)^4 R^2} \int d\Omega_1 \int d\Omega_2 \int
d^3q_1 \int d^3q_2 \int d\omega_1 \\ \nonumber
 \int d\omega_2 Y_{l,m}^*(\Omega_1) Y_{l,-m}^*(\Omega_2)
e^{i \vec{q}_1\cdot\hat{\Omega}_1 R}
  e^{i \vec{q}_2\cdot\hat{\Omega}_2 R} \hat{\Omega}_1^\alpha
\hat{\Omega}_2^\beta
 \Big\langle G^i(\vec{q}_1,\omega-\omega_1)\\ \nonumber
  G^i(\vec{q}_2,-\tilde{\omega}-\omega_2)  \Big(
  v_{ext}^\alpha(\vq_1,\go_1)  + v_r^{\alpha}(\vq_1,\go_1) \Big)
   \Big( v_{ext}^\beta(\vq_2,\go_2) + v_r^{\beta}(\vq_2,\go_2) \Big) \Big\rangle.
\eeq
In this and in all the following equations, the Einstein summation convention is applied to the Cartesian components $\alpha,\beta$. In addition, $G^i(\vq,\go)$ appearing in the equation above, is defined as the temporal FT of
\beq \label{eq:G} G^i(\vec{q},t) = \exp(i\vec{q}\cdot
\vec{r}_0^i(t)). \eeq
%
%
The first order correction in the density has two contributions. The
first contribution is obtained by neglecting $\vvi$ in both brackets on
the right hand side of eq. (\ref{eq:37}) but taking the $G$'s to
first order in $n$. This results in a contribution ${\cal G}_{l,1}^{(1)}$
given by
\beq \label{eq:flmflmcor}
n{\cal G}_{l,1}^{(1)}(\go,\tilde{\go})
=-n\delta(\go -\tilde{\go})\frac{1}{\lambda_l^2+\go^2}
(2\pi)^{-2}R^{-2} \int d\gO_1 \int d\gO_2 \int d^3q
Y_{l,m}^*(\gO_1) Y_{l,-m}^*(\gO_2)  \\ \nonumber
e^{i\vq\cdot(\hat{\gO}_1- \hat{\gO}_2)R}  \hat{\gO}_1^\alpha
\hat{\gO}_2^\beta  (\delta_{\alpha,\beta} - \frac{q^\alpha
   q^\beta}{q^2}) \int dt
 e^{-i\go t}e^{-\frac{q^2}{6}{\cal F}_0(t)}\phi(q,t)\frac{q^2}{6}{\cal
   F}_1(t).
\eeq
%
In this general form ${\cal F}={\cal F}_0 + {\cal F}_1 \cdot n$ is the total mean square displacement (MSD) of the center of mass. The above contribution vanishes, however, for cases
where the bare velocity field is uncorrelated in time, eq. (\ref{eq:PhiWhite}).
Consider the integral over $t$ on the right hand side of eq. (\ref{eq:PhiWhite}). The Dirac delta function in $\phi(q,t)$, eq. (\ref{eq:PhiWhite}), sets $t=0$. Since for $t=0$ the MSD must vanish in any order of the expansion in $n$, ${\cal F}_1(0)=0$ and the right hand side of eq. (\ref{eq:flmflmcor}) vanishes.\\
The second contribution arises from those terms in (\ref{eq:37}) linear in
$\vvi$, taking the MSD to zero order in $n$. In the linear approximation and for small deformations, the velocity
field created by the  deformable objects can be written as a sum of
prefactors, $O_{l,m}^j$ that multiply the deformation modes of the
$j$'th object. Considering a set of identical objects,
it is obvious that $O_{l,m}^j$ is identical for all $j$. Thus,
the superscript is dropped and $\vvi$ is written as
\beq \label{eq:vi26}
\vvi(\vec{r},t)=\sum_{j \neq i} \sum_{l,m}
  O_{l,m}(\vec{r}-\vec{r}_0^j(t))f_{l,m}^j(t).
\eeq
The prefactors $O_{l,m}$ are model dependent and are calculated in appendix \ref{sec:velocity}
for deformable objects that are governed by surface tension.
Using the above expression for $\vvi$, the second contribution is given by
\beq \label{eq:calg2}
{\cal G}_{l,1}^{(2)}(\go,\tilde{\go})\equiv
-2{\cal R}\Bigg\{
\delta(\go-\tilde{\go}) \frac{1}{\lambda_l^2 + \go^2}
\frac{1}{(2\pi)^{\frac{9}{2}} R^3} \int d\gO_1 \int
d\gO_2 \int d^3q_1 \int d^3q_2\\ \nonumber
Y_{l,m}^*(\gO_1)
Y_{l,-m}^*(\gO_2)
e^{i\vq_1\cdot\hat{\gO}_1 R}
e^{i\vq_2\cdot \tilde{\gO}_2 R} \hat{\gO}_1^\alpha
\hat{\gO}_2^\beta
\sum_{l',m'} O_{l',m'}^\alpha(\vq_1) 
 \int d\gO_3 
Y_{l',m'}^*(\gO_3)\\ \nonumber
 e^{-i\vq_2 \cdot\hat{\gO}_3R}
\hat{\gO}_3^\gamma 
(\delta_{\gamma,\beta} - \frac{q_2^\gamma   q_2^\beta}{q_2^2})
%
%
\frac{\tilde{\phi}(q_2)}{i\go+\lambda_{l'}+q_1^2D}
\Big( S_{\vq_1+\vq_2} -1 \Big)\Bigg\},
\eeq
where ${\cal R}\{ x\}$ denote the real part of $x$ 
(for detailed derivation of equation (\ref{eq:calg2}) see appendix \ref{app:B}).

Once $O_{l,m}$ and $\lambda_l$
\cite{schwartz88,safran87,foltin,gang95,komura} of a
specific system and the correlations of the bare velocity field are known, the
correction to the shape correlations can be calculated using the above
equations. In what follows, the actual use of the general
equations is demonstrated for a specific system of deformable
objects that are governed by surface tension.

\section{Droplets Governed by Surface Tension} \label{sec:white}
To determine the shape correlations we need $\tilde{\phi}(q)$ that describes the effect of external agents on the system. We need the set of the $\lambda_l$'s describing the decay of a deformation of angular momentum $l$ of a given
membrane in the absence of the bare velocity field. We also need
the $O_{l,m}$'s that describe how the velocity field in the liquid
responds to the deformation of a single object. The above quantities depend on the properties of the deformable objects and of the liquid in which they are immersed. \\
The system to be considered in the following is a system of membranes governed by surface tension. Namely, the energy of the membrane $U_S$ is given by 
$U_S=\lambda S$ where $S$ is the surface area. The
viscosity, $\eta$, is assumed to be uniform inside and outside the objects.
(The qualitative behaviour for different viscosities is not changed
for a finite reasonable range of viscosity ratios
\cite{hinch} and using identical viscosities eliminates the boundary
conditions, thus simplifying the calculations considerably).
For that case the $\lambda_l$'s were derived in the past
\cite{schwartz88},
\beq
\lambda_l=\frac{\lambda}{4\eta
  R}\frac{(l+2)(l+1)l(l-1)}{(l+\frac{3}{2})(l+\frac{1}{2})(l-\frac{1}{2})}.
\eeq

The $O_{l,m}$'s, for the same system, are needed only for large
distances from the centre of the object inducing the velocity by its
deformations, because the density of deformable objects is low. The
leading non trivial contribution to the $O_{l,m}$ is calculated in
appendix \ref{sec:velocity} and is given by
\beq \label{eq:olmst}
O_{l,m}(\vec{r})=\delta_{l,2} \frac{\lambda}{\eta}\left(\frac{R}{r}\right)^2
\frac{4}{5}  Y_{2,m}(\gO) \hat{\gO},
\eeq
where $\hat{\gO}$ is the unit vector in the direction of $\vec{r}$.

Now, the integration over $\gO_1$, $\gO_2$ and $\gO_3$ in eq. (\ref{eq:calg2}) can be easily
done using the partial wave expansion,
\begin{eqnarray} \label{eq:partial_waves}
e^{-i\vec{q}\cdot(R\hat{\gO})} = \sum_{l=0}^{\infty} \sum_{m=-l}^{l}
(-i)^l 4\pi j_l(qR)Y_{lm}^*(\Omega_q)Y_{lm}(\Omega).
\end{eqnarray}
These integrations produce a long but finite set of terms that is not presented here because of its length and complexity.
To facilitate the integration over the $q$'s, the following tactics is used.
The above expression is split into three parts by defining the following Fourier
transforms:

\beq
A_{m,m'}(\vec{r}_1)= \int d\gO_1 \int d^3q_1 \frac{e^{i\vq_1\cdot\hat{\gO}_1
    R}}{i\go +\lambda_2 + q_1^2D} O_{2,m'}^\alpha(\vq_1)
Y_{l,m}^*(\gO_1)\hat{\gO}_1^\alpha \frac{e^{i\vq_1\cdot\vec{r}_1}}{(2\pi)^{\frac{3}{2}}},
\eeq
\beq
B_{m,m'}(\vec{r}_2)= \int d\gO_2 \int d\gO_3 \int d^3q_2
e^{i\vq_2\cdot(\hat{\gO}_2 + \hat{\gO}_3)R} (\delta_{\gamma,\beta} - \frac{q_2^\gamma
 q_2^\beta}{q_2^2}) \tilde{\phi}(q_2) \\ \nonumber
Y_{l,-m}^*(\gO_2)Y_{2,m'}^*(\gO_3) \hat{\gO}_2^\beta \hat{\gO}_3^\gamma   \frac{e^{i\vq_2\cdot\vec{r}_2}}{(2\pi)^{\frac{3}{2}}},
\eeq
and
\beq
C(\vec{r}_3) = \int d^3q_3 \Big( S_{\vq_3} -1 \Big) \frac{e^{i\vq_3\cdot\vec{r}_3}}{(2\pi)^{\frac{3}{2}}}.
\eeq

Combining the above together, the first correction to the shape
correlation for a dilute system is obtained by
%
%
changing the order of integration and performing first the integration
over the $q$'s,

\beq
n{\cal G}_{l,1}(\go,\tilde{\go})=
2{\cal R}\Bigg\{
\frac{ \delta(\go-\tilde{\go})}{(\lambda_2^2 + \go^2)R^3
  (2\pi)^2} \int d^3r \sum_{m'}
A_{m,m'}(\vec{r})B_{m,m'}(\vec{r})C(-\vec{r})\Bigg\}.
\eeq

Note that since the system is supposed to be invariant
under rotations, the structure factor can depend only on the absolute
value of $q$ and $C$ must depend only on the absolute value of $r$.

An additional property of the shape correlations is that the
expression:\\
$ \frac{1}{2l+1}\sum_{m} \left\langle f_{l,m}^i(\omega)
  {f_{l,m}^i}^*(-\tilde{\omega})\right\rangle$
transforms as a scalar. Moreover, due to the rotational symmetry, the
shape correlation does not depend on $m$ and therefore,
\beq
 \left\langle f_{l,m}^i(\omega)
   {f_{l,m}^i}^*(-\tilde{\omega})\right\rangle= \frac{1}{2l+1}\sum_{m'}
  \left\langle f_{l,m'}^i(\omega)  {f_{l,m'}^i}^*(-\tilde{\omega})\right\rangle.
\eeq
In this way the coordinate system is easily rotated without loss of
generality.

Last, note that the expression for $C(r)$ is, up to a prefactor of
$n(2\pi)^{-\frac{3}{2}}$, no other than the
pair distribution function.
The pair distribution function $C$ is chosen,
as a good approximation in the dilute regime,
to be the pair distribution function of hard spheres system in the
dilute limit,
\beq
C(r)=\left\{ \begin{array}{ll}
0 & $if $ r<2R\\
n^2 & $if $ r>2R.
\end{array} \right.
\eeq

A straightforward but tedious derivation, shows that the
shape correlations for all $l$ except $l=2$ vanish
 and for $l=2$ it is given by the following expression:

\beq \label{eq:flmlast}
{\cal G}_{l,1}(\go,\tilde{\go})=-\Re\Bigg\{\delta_{l,2}
\frac{512\sqrt{2}\delta(\go-\tilde{\go})}{3\sqrt{\pi}(\lambda_2^2 +
  \go^2)} \frac{\lambda}{\eta R^{13}}
\int_{2R}^\infty dr \int_0^\infty dq
\quad \quad \quad \quad \quad \quad \quad
\\ \nonumber
\frac{(3\cos(qR)qR-3\sin(qR)+\sin(qR)q^2R^2)}{(i\go+\lambda_2+q^2D)q^4}
\int_0^\infty dq_2
\tilde{\phi}(q_2)\sin(qr)\sin(q_2r)\\ \nonumber
\frac{\left(-9\cos(q_2R)q_2R+9\sin(q_2R)-4\sin(q_2R)q_2^2R^2+\cos(q_2R)q_2^3R^3\right)^2}
{q_2^7}\Bigg\}.
\eeq
%

\begin{figure}[!htp]
\centerline{\psfig{figure=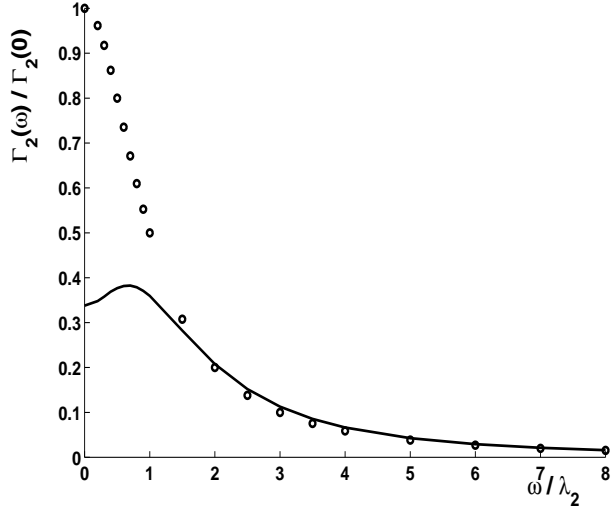,width=9cm ,height=7cm}}
\caption{The correlations of the deformation coefficient with
$l=2$, $\langle f^i_{2,m}(\go) {f^i_{2,m}(-\tilde{\go})}^*
\rangle$, to zeroth order (dotted line) and first order
(continuous line) in the density of objects. The units along the y
axis are relative and the units along the x axis $\go$ are
normalized by the decay rate $\lambda_2$.} \label{fig1}
\end{figure}

The integration over $q$ can be done analytically but will not be
presented here due to the length of the expression.
Note also that the integration over $r$ must be done last.
Once $\tilde{\phi}$ is known, the result can be calculated
analytically for special cases or numerically for others.
%
A specific example for the use of eq. (\ref{eq:flmlast}) is depicted
in fig. \ref{fig1}, for a correlation function that has the form of the Yukawa
potential. The Fourier transform of the correlation function is given by
\beq
\tilde{\phi}(q_2)=\phi_0/(a^2+q_2^2),
\eeq
where $a=\frac{1}{R^2}$ is used in this example.
In addition 
$\frac{\lambda_2  R^2}{D}=1$ is used.
The correlation of the deformation
coefficients, $f_{l,m}^i$, with $l=2$ is depicted to zeroth
order (dotted line) and first order (continuous line) in the density
of objects.
The correlation of the deformation coefficients must have the form,
\beq
\left\langle f_{l,m}^i(\go) {f_{l,m}^i}^*(-\tilde{\go})\right\rangle =
\delta(\go-\tilde{\go}) \Gamma_l(\go),
\eeq
where $\Gamma$ is a general function that depends on $l$ and $\go$.
In general, the correction that
is linear in density, $n{\cal G}_1$, must be small relatively to the
zeroth order term ${\cal G}_0$. In this example however, the density $n$ is chosen to be large enough to observe changes. 
The correlations of deformation coefficients with $l \neq 2$ do not
change to first order in the density of the deformable objects.
As can be seen, the deformation modes with $l=2$ are suppressed by the 
interaction between the droplets at low frequencies, $\omega<\lambda_2$. 
This is expected due to the retarded response of each droplet to the external velocity field. This decay rate $\lambda_2$ introduces a new time scale that controls the decay of fluctuations produced by the external field. At low frequencies, lower than the time it takes the deformation modes to decay; the velocity field, induced by neighbouring droplets, responds effectively to the external velocity field and thus decreases the $f_{l,m}$ correlation at low frequencies.

\appendix
\section{On the calculation of the deformation correlations} \label{app:B}
This appendix derives the contribution of the terms involving
$v_{ext}^{\alpha}(\vq,\go_1) v_r^{\beta}(\vq_2,\go_2)$ and
$v_r^{\alpha}(\vq,\go_1) v_{ext}^{\beta}(\vq_2,\go_2)$ in
eq. (\ref{eq:37}) (i.e. ${\cal G}_{l,1}^{(2)}$).
These two terms are complex conjugates of each other. Thus only the
first term will be considered and the correction ${\cal G}_{l,1}^{(2)}$
is given by twice the real part of the answer.
In order to keep the expressions to first order in the deformation and
density of objects, expressions must be kept linear in $\vvi$ and
$f_{l,m}$.
First, the average is broken into two parts,
the velocity correlation and the average
over expressions containing $\vro$ \cite{gady2} ( This approximation was justified and used a number of times in the past \cite{gady2,brus91}).

\beq
{\cal G}_{l,1}^{(2)}(\go,\tilde{\go})=
-\frac{1}{(\lambda_l+i\go)(\lambda_l-i\tilde{\go})R^2}(2\pi)^{-7} \int
d\gO \int d\gO_2 \int d^3q \int d^3q_2  \\ \nonumber
\int d\go_1 \int d\go_2 Y_{l,m}^*(\gO) Y_{l,-m}^*(\gO_2) e^{i\vq\cdot\hat{e}_\gO R}
e^{i\vq_2\cdot \tilde{e}_{\gO_2}R} \hat{e}_{\gO}^\alpha
\hat{e}_{\gO_2}^\beta \Big\langle G^i(\vq,\go-\go_1)
G^i(\vq_2,-\tilde{\go}-\go_2)
  \\ \nonumber
\sum_{j\neq i}\sum_{l',m'} O_{l',m'}^\alpha(\vq)
\int dt e^{-i\go_1 t} e^{-i\vq\cdot \vro^j(t)} \int d\go_3 e^{i\go_3 t}
\frac{1}{(\lambda_{l'}+i\go_3)R} \int d\gO_3 \int d^3q_3
Y_{l',m'}^*(\gO_3)\\ \nonumber
 e^{i\vq_3 \cdot\hat{e}_{\gO_3}R}
\hat{e}_{\gO_3}^\gamma \int d\go_4
G^j(\vq_3,\go_3-\go_4) \Big\rangle
\Big\langle \vve^{\gamma}(\vq_3,\go_4) \vve^{\beta}(\vq_2,\go_2)
\Big\rangle
\eeq

The Use of the expression for the correlation of the bare velocity
,eq. (\ref{eq:fcorelation}), 
and the definition of $G^i$, eq (\ref{eq:G}), yield

\beq
{\cal G}_{l,1}^{(2)}(\go,\tilde{\go})=
-\frac{1}{(\lambda_l+i\go)(\lambda_l-i\tilde{\go})R^2}(2\pi)^{-6} \int
d\gO \int d\gO_2 \int d^3q \int d^3q_2  \\ \nonumber
 \int d\go_1 \int d\go_2
Y_{l,m}^*(\gO) Y_{l,-m}^*(\gO_2) e^{i\vq\cdot\hat{e}_\gO R}
e^{i\vq_2\cdot \tilde{e}_{\gO_2}R} \hat{e}_{\gO}^\alpha
\hat{e}_{\gO_2}^\beta \sum_{j\neq i} \sum_{l',m'} O_{l',m'}^\alpha(\vq) \\ \nonumber
\int d\go_3 \frac{1}{(\lambda_{l'}+i\go_3)R}  \int d\gO_3
Y_{l',m'}^*(\gO_3) e^{-i\vq_2 \cdot\hat{e}_{\gO_3}R}
\hat{e}_{\gO_3}^\gamma
(\delta_{\gamma,\beta} - \frac{q_2^\gamma
   q_2^\beta}{q_2^2})\phi(q_2,\go_2) \\ \nonumber
\Big\langle G^i(\vq,\go-\go_1) G^i(\vq_2,-\tilde{\go}-\go_2)
G^j(-\vq,\go_1-\go_3)G^j(-\vq_2,\go_3+\go_2) \Big\rangle
\eeq

where
\beq
\Big\langle G^i(\vq,\go-\go_1) G^i(\vq_2,-\tilde{\go}-\go_2)
G^j(-\vq,\go_1-\go_3)G^j(-\vq_2,\go_3+\go_2) \Big\rangle =\\ \nonumber
\frac{1}{(2\pi)^2}
\int dt_1
e^{-i(\go-\go_1)t_1}
 \int dt_2 e^{i(\tilde{\go}+\go_2)t_2} \int dt_3 e^{-i(\go_1-\go_3)t_3} \int
  dt_4 e^{-i(\go_3+\go_2)t_4}\\ \nonumber
\Big\langle \exp(i\vq\cdot(\vro^i(t_1)-\vro^j(t_3)))
\exp(i\vq_2\cdot(\vro^i(t_2)-\vro^j(t_4))) \Big\rangle.
\eeq
The integration over $\go_1$ can be done using $\int d\go_1
e^{i\go_1(t_1-t_3)}=2\pi \delta(t_1-t_3)$.
Decoupling the average into a product of averages one finds that
\beq
\sum_{j\neq i}\Big\langle \exp\Big(i\vq\cdot(\vro^i(t_1)-\vro^j(t_1))\Big)
\exp\Big(i\vq_2\cdot(\vro^i(t_2)-\vro^j(t_4))\Big) \Big\rangle \\ \nonumber
=\sum_{j \neq i}
\left\langle \exp\Big(i\vq\cdot(\vro^i(t_1)-\vro^i(t_2))\Big)\right\rangle
\left\langle
  \exp\Big(-i\vq\cdot(\vro^j(t_1)-\vro^j(t_4))\Big)\right\rangle\\ \nonumber
\left\langle \exp\Big(i(\vq+\vq_2)\cdot(\vro^i(t_2)-\vro^i(t_4))\Big)\right\rangle
\left\langle \exp\Big(-i(\vq+\vq_2)\cdot(\vro^j(t_4)-\vro^i(t_4))\Big)\right\rangle\\ \nonumber
=
\exp\Big(-\frac{q^2}{6}{\cal F}(t_1-t_2)
-\frac{q^2}{6}{\cal F}(t_1-t_4)
-\frac{(\vq+\vq_2)^2}{6}{\cal F}(t_2-t_4)\Big)\\ \nonumber
\frac{1}{N}\sum_i \sum_{j \neq i}\left\langle
  \exp\Big(i(\vq+\vq_2)\cdot\vro^i(t_4)\Big)
  \exp\Big(-i(\vq+\vq_2)\cdot\vro^j(t_4)\Big)\right\rangle,
\eeq
where $N$ is the number of deformable objects in the system.
The reason for the decoupling of exponents is that the driving external velocity correlation decays both in time and with distance. Because of the low density and the short range repulsion (as of hard spheres in our approximate description to the first order in the density of objects) the exponents can be only weakly correlated.\\
The addition and subtraction of the $i$'th object to the sum over $j$
results in
\beq
&&\sum_{j\neq i}\Big\langle \exp(i\vq\cdot(\vro^i(t_1)-\vro^j(t_1)))
\exp(i\vq_2\cdot(\vro^i(t_2)-\vro^j(t_4))) \Big\rangle=\\ \nonumber
&&e^{-\frac{q^2}{6}{\cal F}(t_1-t_2)}
e^{-\frac{q^2}{6}{\cal F}(t_1-t_4)}
e^{-\frac{(\vq+\vq_2)^2}{6}{\cal F}(t_2-t_4)}
\Big( S_{\vq+\vq_2} -1 \Big).
\eeq

By combining all the above together, using variable transformation and performing
the integrations over $\go_3$, $\go_2$ and $t_4$, the shape
correlation are derived,

\beq \label{eq:flmei}
{\cal G}_{l,1}^{(2)}(\go,\tilde{\go})=
-2{\cal R}\Bigg\{\delta(\go-\tilde{\go}) \frac{1}{\lambda_l^2 + \go^2} \frac{1}{(2\pi)^{\frac{9}{2}} R^3} \int d\gO \int
d\gO_2 \int d^3q \int d^3q_2  \\ \nonumber
Y_{l,m}^*(\gO) Y_{l,-m}^*(\gO_2) e^{i\vq\cdot\hat{e}_\gO R}
e^{i\vq_2\cdot \tilde{e}_{\gO_2}R} \hat{e}_{\gO}^\alpha
\hat{e}_{\gO_2}^\beta
\sum_{l',m'} O_{l',m'}^\alpha(\vq)
 \int d\gO_3
Y_{l',m'}^*(\gO_3) e^{-i\vq_2 \cdot\hat{e}_{\gO_3}R}
\hat{e}_{\gO_3}^\gamma  \\ \nonumber
(\delta_{\gamma,\beta} - \frac{q_2^\gamma   q_2^\beta}{q_2^2})
\int_0^\infty dt_1 \int_{-\infty}^\infty dt_2 e^{-i\go(t_1-t_2)}
\phi(q_2,t_2) e^{-\lambda_{l'}t_1}
e^{-\frac{q^2}{6}{\cal F}(t_1-t_2)}
e^{-\frac{q^2}{6}{\cal F}(t_1)}
\\ \nonumber
e^{-\frac{(\vq+\vq_2)^2}{6}{\cal F}(t_2)}
\Big( S_{\vq+\vq_2} -1 \Big)\Bigg\},
\eeq

from which (\ref{eq:calg2}) is obtained.

\section{the velocity field generated by deformation for the surface
  tension case} \label{sec:velocity}

Consider a liquid droplet governed by surface tension that is
immersed in a host liquid. Assume that the viscosities inside the
droplet and in the host liquid are equal. The deformation of the shape of
the object changes its energy and in response induces a force density
that acts on the fluid. The force density creates in its turn an
additional velocity field, denoted here as $\vec{v}_\psi$,

\beq \label{eq:oseen1}
\vec{v}_\psi(\vec{r})=\frac{1}{\eta}\int
S(\vec{r}-\vec{r}')\cdot\vec{F}(\vec{r}')d^3r',
\eeq

where $\vec{F}$ is the force density created by the object and $S$ is the Oseen tensor that is given by
\beq
S_{i,j}(\vec{r})=\frac{1}{8\pi}\left( \frac{\delta_{i,j}}{r}+\frac{r_i  r_j}{r^3}\right)
\eeq

It is easy to calculate the force density using simple field theory.
Let $\psi(\vec{r})$ be a three dimensional scalar field, defined
everywhere in such a way that the equation $\psi(\vec{r})=0$ describes the surface of the object \cite{schwartz88,schwartz90b}. The gradient of $\psi$ is assumed
to exist and not to vanish in the vicinity of $\psi(\vec{r})=0$.
Under the additional assupmtion that the deformation of the object do not produce over-hangs,$\psi(\vec{r})$ is written using the deformation function, $f(\gO)$, given in
eq. (\ref{eq:fOmegat}).
\beq \label{eq:psif}
\psi=\frac{r}{R}+f(\gO,t)-1.
\eeq

In \cite{schwartz88} the force density created by a deformed objects
governed by surface tension is given by,
\beq
\vec{F}(\vec{r})=-\lambda (\divg\hat{n})\delta(\psi(\vec{r}))\grad \psi(\vec{r}),
\eeq

where $\hat{n}=\frac{\grad\Psi}{|\grad\psi|}$ is a unit vector in the
direction normal to the surface of the deformable object.

In addition, the following symbols are used for the angular parts of the
gradient and Laplacian,

\beq
\D=\hat{\theta}\frac{\partial}{\partial\theta}+\hat{\varphi}\frac{1}{\sin(\theta)}\frac{\partial}{\partial\varphi}
\eeq
and
\beq
\D^2=\frac{1}{\sin(\theta)}\frac{\partial}{\partial\theta}\left(\sin(\theta)\frac{\partial}{\partial\theta}\right)
+\frac{1}{\sin^2(\theta)}\frac{\partial^2}{\partial\varphi^2},
\eeq
where $\hat{\theta}$ and $\hat{\varphi}$ are unit vectors of $\theta$ and $\varphi$.
The velocity field $\vec{v}_{\psi}$ is obtained as a function
of $f(\gO',t)$ by the use of equations (\ref{eq:oseen1}) and (\ref{eq:psif}),
\beq
\vec{v}_{\psi}(\vec{r})=-\frac{\lambda}{\eta}\int
S(\vec{r}- R(1-f)\hat{\gO}')\Bigg[ R \Bigg( \frac{
  \frac{2}{1-f}+\frac{\D^2f}{(1-f)^2}}{\sqrt{
    1+\frac{(\D f)^2}{(1-f)^2}}} \\ \nonumber
-\frac{1}{2}\frac{\hat{\gO}'+\frac{\D
    f}{1-f}}{(1+\frac{(\D f)^2}{(1-f)^2})^{\frac{3}{2}}}\cdot \Big( -2 \frac{(\D f)^2}{(1-f)^3}\hat{\gO}' + \frac{\D((\D f)^2)}{(1-f)^3}\Big)\Bigg)
\Big(\hat{\gO}'+\frac{\D f}{1-f}\Big)\Bigg] (1-f)^2 d\gO'.
\eeq

Keeping the above expression to the first order of the deformation $f$
results with

\beq \label{eq:vpsi2}
\vec{v}_{\psi}(\vec{r})=\frac{\lambda R}{8\pi \eta} \int d\gO'
\Bigg[& &
\Big(
\frac{2\hat{\gO}'}{X}+2\frac{\vec{X}\cdot\hat{\gO}'}{X^3}\vec{X}-\frac{2}{X^3}R\vec{X}+6\frac{(\vec{X}\cdot\hat{\gO}')^2}{X^5}R\vec{X}\Big)f(\gO')\\ \nonumber
&-& \Big(\frac{2\D f(\gO')}{X}+\frac{2 \vec{X}\cdot(\D f(\gO'))}{X^3}\vec{X}\Big) \\ \nonumber
&-&\Big(
\frac{\hat{\gO}'}{X}+\frac{\vec{X}\cdot\hat{\gO}'}{X^3}\vec{X}\Big)\D^2 f(\gO')
\Bigg],
\eeq
where $\vec{X} \equiv \vec{r} - R\hat{\gO}'$.
$f$ is replaced by its expansion in spherical harmonics
$f(\gO,t)=\sum_{l=2}^{\infty} \sum_{m=-l}^{l} f_{l,m}(t)Y_{l,m}(\gO)$.
The spherical harmonics are eigenvalues of $\D^2$.
In contrast, $\D$ mixes different harmonics.
The density of objects is assumed to be low and therefore the distance
$r$ to the closest droplet is typically much larger than the radius of
the droplet $R$. Hence, the velocity is expanded to the first nontrivial
order in $\frac{R}{r}$.
In addition, a rotated coordinate system is used in which
$\vec{r}=r\hat{z}'$, where $\hat{z}'$ is the $z$ direction in the
rotated coordinate system. In that system,
\beq
\vec{v}_\psi(r\hat{z}',t)=\frac{\lambda}{\eta}\left(
  \frac{R}{r}\right)^2 \frac{2}{\sqrt{5\pi}}f'_{2,0}\hat{z}',
\eeq
where $f'_{2,0}$ is the deformation coefficient of
$Y'_{2,0}$ in the rotated coordinate system.
The velocity field in a general direction is given by a
rotation of the coordinate system. The
transformation under rotation of the spherical harmonics (and thus of $f_{2,0}$)
is given by the addition theorem,
\beq
P_n(\cos(\gamma))=\frac{4\pi}{2n+1}\sum_{m=-n}^{n} Y_{n,m}(\theta_1,\varphi_1)Y_{n,m}^*(\theta_2,\varphi_2),
\eeq
where $\gamma$ is the angle between the directions $\hat{\gO}_1$ and
$\hat{\gO}_2$ that corresponds to $(\theta_1,\varphi_1)$ and
$(\theta_2,\varphi_2)$.
Note that the calculation of $f'_{2,0}$ involves integration over all
angles that produces a dependence of the induced velocity on the
angles at the original coordinate system.
Thus the induced velocity in a general
direction is given by,

\beq \label{eq:vpsir}
\vec{v}_\psi(\vec{r})=\frac{\lambda}{\eta}\left(\frac{R}{r}\right)^2
\frac{4}{5} \sum_{m=-2}^{2} Y_{2,m}(\gO)f_{2,m}(t) \hat{r},
\eeq
where $\hat{r}$ is a unit vector in the
direction of $\vec{r}$. $O_{l,m}$ is obtained by comparing
eq. (\ref{eq:vi26}) with eq. (\ref{eq:vpsir}),
\beq
O_{l,m}=\delta_{l,2} \frac{\lambda}{\eta}\left(\frac{R}{r}\right)^2
\frac{4}{5}  Y_{2,m}(\gO) \hat{r}.
\eeq

Hence, the only spherical harmonics modes that contribute to the
velocity field, far from the object, are the terms with $l=2$.

%
%
%

%
%
%
%
%
%

%
%
%
%
%
\end{document}